\documentclass[seceq]{ptptex}

\usepackage{graphicx}
\usepackage{wrapft}

\newcommand {\tr}{{\rm tr\,}}

\newcommand {\beq}{\begin{equation}}
\newcommand {\eeq}{\end{equation}}
\newcommand {\beqa}{\begin{eqnarray}}
\newcommand {\eeqa}{\end{eqnarray}}

\newcommand {\ee}{\mbox{e}}

\newcommand\fverb{\setbox\pippobox=\hbox\bgroup\verb}
\newcommand\fverbdo{\egroup\medskip\noindent%
                        \fbox{\unhbox\pippobox}\ }
\newcommand\fverbit{\egroup\item[\fbox{\unhbox\pippobox}]}
\newbox\pippobox

\title{
A non-perturbative study of
non-commutative U(1) gauge theory
\footnote{Talk presented by J.\ Nishimura at the
21st Nishinomiya-Yukawa Memorial Symposium on Theoretical Physics: 
``Noncommutative Geometry and Quantum Spacetime in Physics'',
Nishinomiya and Kyoto (2006). \
Preprint KEK-TH-1162, DESY-07-091, HU-EP-07/20}
}

\author{
Jun \textsc{Nishimura}$^{\rm \, a,b}$, 
Wolfgang \textsc{Bietenholz}$^{\rm \, c}$,
Yoshiaki \textsc{Susaki}$^{\rm \, a,d}$,
Jan \textsc{Volkholz}$^{\rm \, e}$ %
}

\inst{
$^{\rm a \,}$ High Energy Accelerator Research Organization (KEK), \\
Tsukuba, Ibaraki, 305-0801, Japan \\
\vspace*{2mm}
$^{\rm b \,}$ Department of Particle and Nuclear Physics,\\
Graduate University for Advanced Studies (SOKENDAI),\\
Tsukuba, Ibaraki 305-0801, Japan \\
\vspace*{2mm}
$^{\rm c}$ John von Neumann Institut f\"{u}r Computing (NIC) \\ 
Deutsches Elektron Sychrotron (DESY) \\
Platanenallee 6, D-15738 Zeuthen, Germany \\
\vspace*{2mm}
$^{\rm d}$ Graduate School of Pure and Applied Science \\
University of Tsukuba, Tsukuba, Ibaraki 305-8571, Japan\\ 
\vspace*{2mm}
$^{\rm e}$ Institut f\"{u}r Physik, Humboldt-Universit\"{a}t zu Berlin \\
Newtonstr.\ 15, D-12489 Berlin, Germany \\

}

\abst{
We study U(1) gauge theory on a 4d non-commutative torus,
where two directions are non-commutative.
Monte Carlo simulations are performed
after mapping the regularized theory 
onto a U($N$) lattice gauge theory in $d=2$.
At intermediate coupling strength,
we find a phase in which open Wilson lines acquire non-zero 
vacuum expectation values,
which implies the spontaneous breakdown of translational invariance.
In this phase, various physical quantities
obey clear scaling behaviors
in the continuum limit with a fixed non-commutativity parameter $\theta$,
which provides evidence for a possible continuum theory.
In the weak coupling symmetric phase, 
the dispersion relation involves a negative IR-singular term, 
which is responsible for the observed phase transition.
}

\begin{document}

\maketitle

\section{Introduction}

Non-commutative (NC) geometry has been studied extensively
as a modification of our notion of space-time
at short distances.
It has recently attracted much attention since gauge theories
on a NC geometry have been shown to appear as
a low energy limit of string theories with a background 
tensor field \cite{String}.
NC geometry also appears naturally in matrix model formulations
of string theory \cite{CDS,AIIKKT}. 
Introducing non-commutativity to the space-time coordinates may change
the infrared dynamics drastically 
at the quantum level due to the so-called UV/IR mixing \cite{rf:MRS}.
The same effect also poses a severe problem 
in the renormalization procedure within perturbation theory
since a new type of IR divergences appears in non-planar diagrams.

The finite lattice formulation \cite{AMNS}
(extending an earlier work \cite{AIIKKT})
regularizes such divergences
as well as the ordinary UV divergences.
It therefore provides a non-perturbative framework to establish
the existence of a consistent field theory on a NC geometry.
In ref.\ \citen{2dU1} a simple field theory ---
2d U(1) gauge theory --- on a NC plane
has been studied by Monte Carlo simulation, and the existence of a
finite continuum limit has been confirmed.
The ultraviolet dynamics is described by the commutative
2d U($\infty$) gauge theory. On the other hand, Wilson loops of
large area (with a variety of shapes) pick up a complex phase linear
in the loop area, in the spirit of the Aharonov-Bohm effect 
\cite{2dU1} (but their expectation values are
shape-dependent, in contrast to the commutative plane \cite{BBT}).

As an interesting physical consequence of the UV/IR mixing
in the case of scalar field theory,
ref.\ \citen{GuSo} predicted the existence of a ``striped phase''
for the NC $\lambda \phi^{4}$ model in dimensions $d > 2$.
In this phase, non-zero Fourier modes of the scalar field
acquire a vacuum expectation value
and break the translational invariance spontaneously.
The existence of this new phase was fully established
by Monte Carlo simulations \cite{Bietenholz:2004xs}.
Such a phase has also been observed in the 2d lattice model,
\cite{AC,Bietenholz:2004xs} but in that case the continuum limit
is still under investigation.
An analogous phase was observed in simulations
involving a fuzzy sphere instead of a NC plane \cite{fuzzy}.

In the case of 4d U(1) NC gauge theories,
the one-loop calculation of the effective action \cite{oneloop}
suggests that the perturbative vacuum is 
unstable against the condensation of the Wilson line.
This causes the spontaneous symmetry breaking (SSB) 
of the translational invariance,
since the open Wilson lines carry non-zero momenta \cite{IIKK}.
Whether a stable {\em non-perturbative} vacuum exists
or not is an interesting question,
which we have addressed in ref.\ \citen{Bietenholz:2006cz}
from first principles using the lattice formulation \cite{AMNS}.

\section{Lattice gauge theory on a NC geometry}
\label{sec:lattice}

NC geometry is characterized by the commutation
relation $[ \hat{x}_\mu , \hat{x}_\nu] = i \, \Theta_{\mu \nu}$
among the space-time coordinates $\hat{x}_\mu$,
where $\Theta_{\mu \nu}$ is the non-commutativity tensor.
Here we consider the 4d Euclidean
space-time $\hat{x}_{\mu}$ ($\mu=1,\cdots , 4$) with the 
non-commutativity introduced only in the $\mu=1,2$ directions, i.e.
\beq
\Theta_{12} = - \Theta_{21} = \theta \ ,
\label{def-Theta}
\eeq
and $\Theta_{\mu\nu} = 0$ otherwise.
Since we have two commutative directions, we may regard one of them
as the Euclidean time.
This allows us to alleviate the well-known problems 
concerning causality and unitarity.

The lattice regularized version of gauge theory
on NC geometry can be defined by an
analog of Wilson's plaquette action \cite{AMNS}
\beq
S= - \beta \sum_{x} \sum_{\mu < \nu}
U_\mu (x) \star U_\nu (x + a \hat{\mu}) \star
U_\mu (x + a \hat{\nu})^\ast \star U_\nu (x)^\ast
 + \mbox{c.c.} \ ,
\label{lat-action}
\eeq
where the symbol $\hat{\mu}$ represents 
a unit vector in the $\mu$-direction
and we have introduced the lattice spacing $a$.
Here the star product (denoted by 
$\star$) encodes the non-commutativity of the space-time.

In order to study the lattice NC theory (\ref{lat-action})
by Monte Carlo simulations,
it is crucial to reformulate
it in terms of matrices \cite{AMNS}.
In the present setup (\ref{def-Theta}), with two NC directions
and two commutative ones, the transcription applies only to
the NC plane whereas the commutative plane remains untouched.
Let us decompose the four-dimensional coordinate as 
$x\equiv (y,z)$, where 
$y\equiv (x_1 , x_2)$
and $z\equiv (x_3 , x_4)$
represent two-dimensional coordinates 
in the NC and in the commutative plane, respectively.
We use a one-to-one map between a field $\varphi(x)$
on the four-dimensional $N \times N \times L \times L$ lattice 
and a $N \times N$ matrix field $\hat{\varphi}(z)$ on a two-dimensional 
$L \times L$ lattice.
This map yields the following correspondence
\beqa
\varphi_1 (y,z) \star \varphi_2 (y,z) \quad &\Longleftrightarrow& \quad
\hat{\varphi}_1(z) \, \hat{\varphi}_2(z)  \ , \\
\varphi (y+a\hat{\mu},z) \quad &\Longleftrightarrow& \quad
\Gamma_\mu \, \hat{\varphi}(z) \, \Gamma_\mu^\dag \ , \\
\frac{1}{N^2} \sum_{y} \varphi (y,z) \quad &\Longleftrightarrow& \quad
\frac{1}{N}\, \tr \hat{\varphi}(z) \ .
\eeqa
The SU($N$) matrices $\Gamma_\mu$ ($\mu = 1, 2$),
which represent a shift in a NC direction,
satisfy the 't Hooft-Weyl algebra
with the matrix size $N$ being odd.
For this construction \cite{2dU1,Bietenholz:2004xs},
the non-commutativity parameter $\theta$
in (\ref{def-Theta}) is given by
\beq
\theta = \frac{1}{\pi} N a^2 \ .
\label{theta-def}
\eeq
Note that the extent in the NC directions
$N a$ goes to infinity
in the continuum limit $a \rightarrow 0$ at fixed $\theta$.

Using this map, the link variables $U_\mu(x)$ are mapped
to a $N\times N$ unitary matrix field $\hat{U}_\mu(z)$ on the two-dimensional
$L\times L$ lattice.
The action (\ref{lat-action}) can be rewritten in terms
of $\hat{U}_\mu(z)$.
By performing a field redefinition 
$V_\mu(z) \equiv  \hat{U}_\mu(z) \Gamma_\mu$
for $\mu=1,2$ and $V_\mu(z) \equiv  \hat{U}_\mu(z)$ otherwise,
we arrive at
\beqa
S &=& S_{\rm NC} + S_{\rm com} + S_{\rm mixed} \ , \nonumber \\
S_{\rm NC} &=& 
- N \beta {\cal Z}_{12} \sum_{z} \, \tr \, 
\Bigr( \,  V_1 (z) \, V_2 (z) \, 
V_1 (z)^\dag \, V_2 (z)^\dag \Bigl)
+ \ \mbox{c.c.}  \ ,  \nonumber \\
S_{\rm com} &=& 
- N \beta  \sum_{z} \, \tr  \, 
\Bigr( \, V_3 (z) \, V_4 (z + a \hat{3}) \, 
V_3 (z + a \hat{4})^\dag \, V_4 (z)^\dag  \Bigl)
 \ + \ \mbox{c.c.}  \ , \nonumber \\
S_{\rm mixed} &=& 
- N \beta \sum_{z} \sum_{\mu=1}^2 \sum_{\nu=3}^4  \, \tr  
\, \Bigr( \,
V_\mu (z) \, V_\nu (z)\, 
V_\mu (z + a \hat{\nu})^\dag \, V_\nu (z)^\dag \Bigl) \ + \ \mbox{c.c.}  \ .
\label{action}
\eeqa
This action possesses the U(1)$^2$ symmetry 
$V_\mu(z) \mapsto \ee^{i \alpha_\mu} V_\mu(z) $
for $\mu=1,2$,
which is related to the translational symmetry 
in the NC directions of the action (\ref{lat-action}).

\section{Phase diagram}
\label{sec:phase}

Let us first 
investigate the phase structure of the lattice model (\ref{action}).
As an order parameter for the spontaneous breaking
of the U(1)$^2$ symmetry, we define 
a gauge invariant operator
\begin{equation}
P_\mu(n) = 
\frac{1}{N L^2} \sum_z
\tr 
\Bigl( V_\mu(z)^n \Bigr)  \ ,
\label{orderparameter}
\end{equation}
for $\mu=1,2$.
This operator corresponds to the open Wilson line
carrying a momentum with the absolute value \cite{AMNS,2dU1}
\beq
p=\frac{2\pi k}{Na} \ , \quad
k=\left\{\begin{array}{cc}
\frac{n}{2}~~~\mbox{for even}~ n \ , \\ 
\frac{n+N}{2}~~\mbox{for odd}~ n  \ .
\end{array}\right.
\label{momentum-polyakov}
\eeq
Since the operator $P_\mu(n)$ with odd $n$ carries a momentum
on the cutoff scale, it does not couple to excitations
that survive in the continuum limit.
Therefore we will focus mainly on the even $n$ case in what follows.

\begin{wrapfigure}{r}{6.6cm}
\centerline{\includegraphics[width=5cm,angle=270]{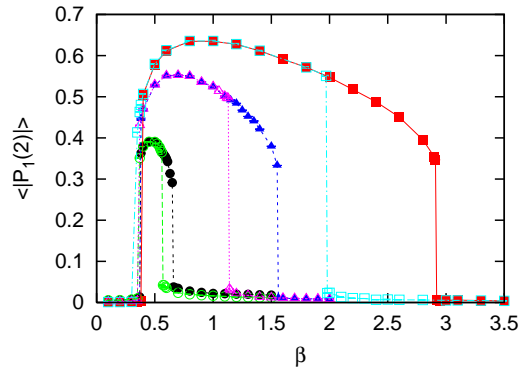}}
\caption{The order parameter $\langle | P_1(n)| \rangle$
is plotted against $\beta$ for $n=2$.
The system size is $N=15$ (circles), $N=25$ (triangles)
and $N=35$ (squares). The closed (open) symbols represent results
obtained with increasing (decreasing) $\beta$, which show
a clear hysteresis behavior.
}
\label{polyakovline}
\end{wrapfigure}

In fig.\ \ref{polyakovline} we plot $\langle | P_1(n)| \rangle$
against $\beta$ for $n=2$.
We observe that there is a phase, in which
the order parameter becomes non-zero.
On the other hand, the quantity $\langle | P_1(n)| \rangle$
for {\em odd} $n$ takes tiny values throughout the whole region of $\beta$.
This implies that the U(1)$^2$ symmetry is broken down 
to $({\rm Z}_2)^2$ in this phase, which we refer to as the ``broken phase''.
The critical point between the broken phase and
the weak coupling phase, denoted as $\beta_{\rm c}$,
increases as $\beta_{\rm c} \sim N^2$. 
    
\section{Continuum limit}
\label{sec:double-scaling}
In this section we investigate whether it is possible
to fine-tune $\beta$ as a function of $N$
in such a way that physical quantities scale.
For that purpose, let us next consider closed Wilson loops, which play
an important role in commutative gauge theories as a criterion
for confinement.
In the present case, since we introduce non-commutativity
only in two directions,
there are three kinds of square-shaped Wilson loops depending on 
their orientations.
For instance, the Wilson loops
in the NC plane and the commutative plane
are defined respectively as
\beqa
W_{12}(n) 
&=&  ({\cal Z}_{12})^{n^2} \, \frac{1}{N L^2} \sum_z
\tr \Bigl( V_1 (z)^n \, V_2 (z)^n \, 
V_1 (z)^{\dag n} \, V_2 (z)^{\dag n} \Bigr)  \ ,  \nonumber \\
W_{34}(n) 
&=& \frac{1}{N L^2} \sum_z  
\tr \Bigl( 
{\cal V}_3 (z , n) \, {\cal V}_4 (z+n a \hat{3}, n)
{\cal V}_3 (z+n a \hat{4},n)^\dag \, {\cal V}_4 (z, n)^\dag \Bigr) \ ,
\label{wilson-def}
\eeqa
using the parallel transporter in the commutative directions
\beq
{\cal V}_\nu (z , n) \equiv
V_\nu (z) V_\nu (z + a \hat{\nu}) \cdots
V_\nu (z + (n-1) a \hat{\nu})  \ .
\eeq
\begin{figure}[htb]
  \parbox{\halftext}{
\centerline{\includegraphics[width=5cm,angle=270]
{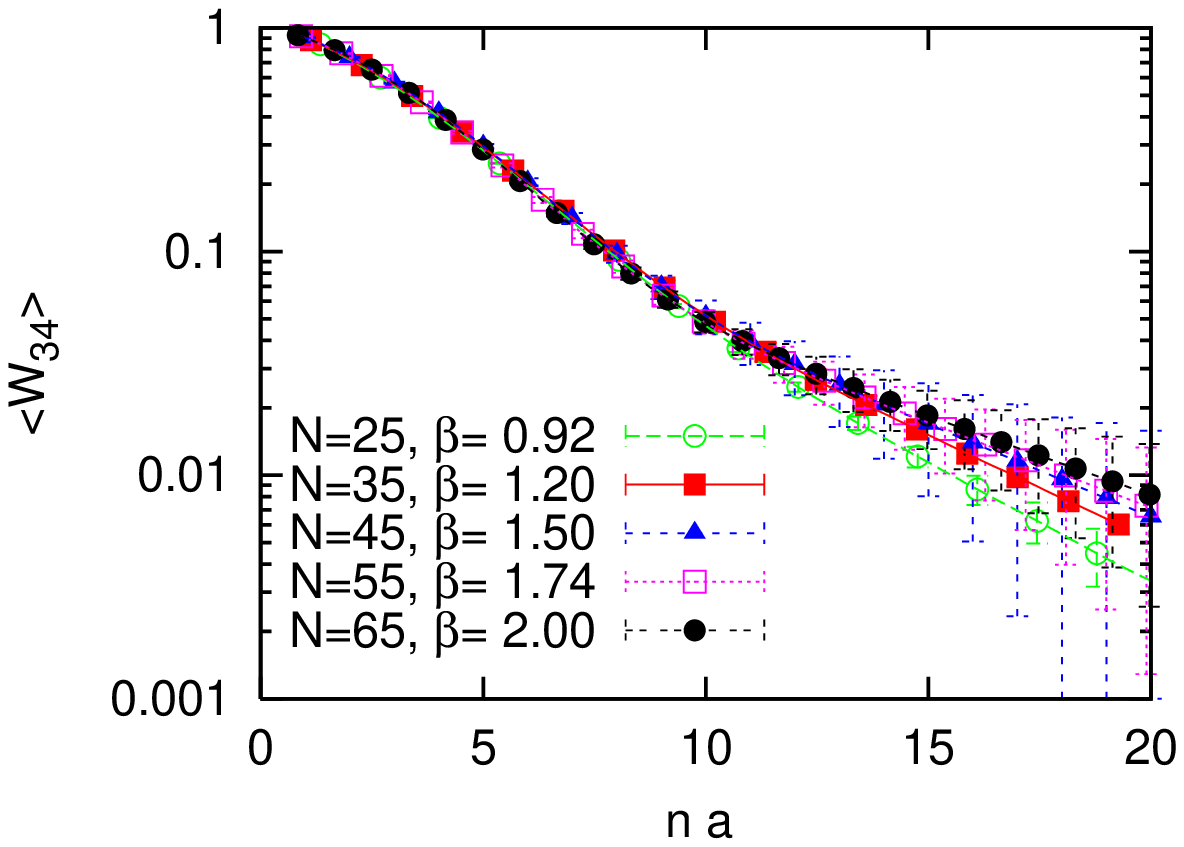}}
\caption{The expectation value of the Wilson loop 
in the commutative plane.}
\label{DSL_Wilsonloops_com}
}
   \hfill
  \parbox{\halftext}{
 \centerline{\includegraphics[width=5cm,angle=270]{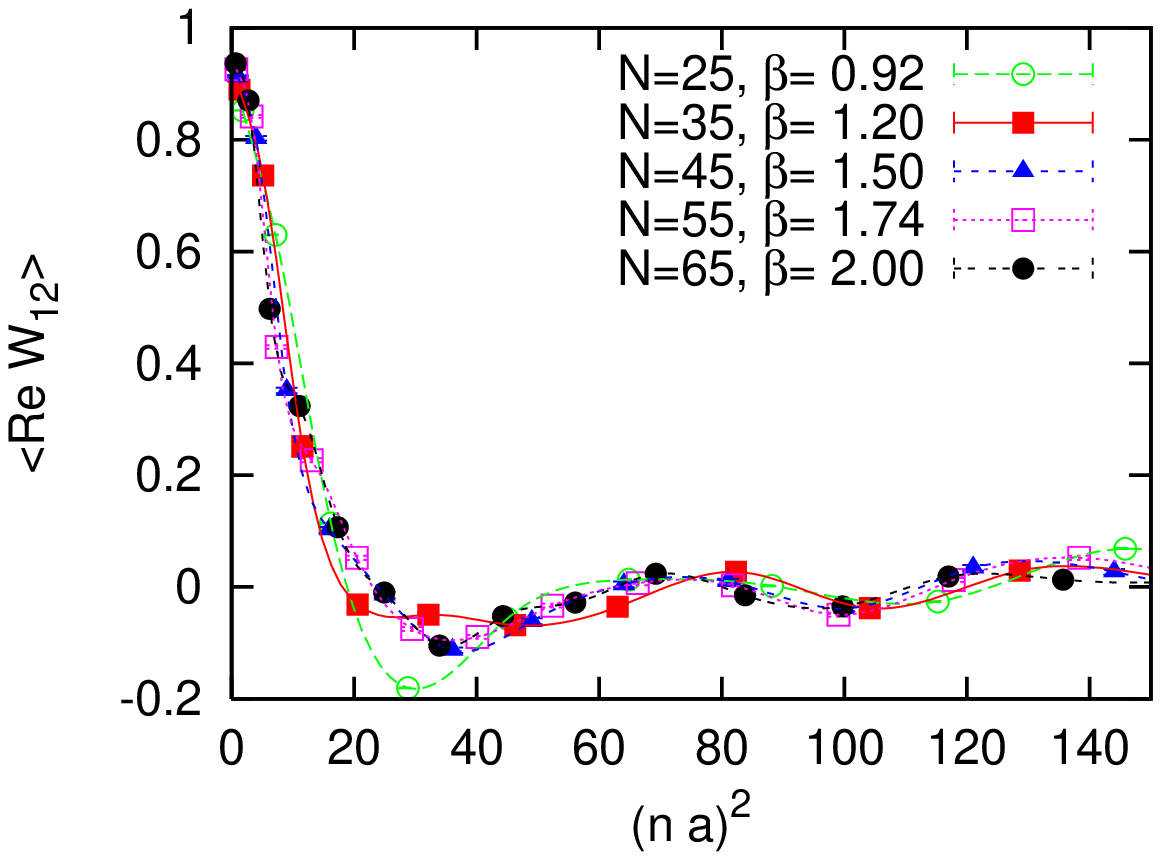}}
      \caption{The expectation value of the Wilson loop 
in the NC plane.}
\label{DSL_Wilsonloops_NC}
}

\end{figure}

In the following we set $a = 1$ for $N=45$ as a convention,
and the lattice spacing $a$ for other $N$ is determined
through (\ref{theta-def}) with $\theta=\frac{45}{\pi}\simeq 14.3$.
As a practical strategy to fine-tune $\beta$, we
first optimize the scaling behavior for
the expectation value of the square Wilson loop $W_{34}(n)$
in the commutative directions.
Fig.\ \ref{DSL_Wilsonloops_com} shows the result.
The horizontal axis represents the physical side length $na$ 
of the loop. We observe a clear scaling behavior,
and the scaling region extends as $N$ increases.
In fact the optimal $\beta$ increases with $N$ much slower than
the lower critical point $\beta_c$ between the 
broken phase and the weak coupling phase, which grows as $N^2$.
This implies that we remain in the broken phase in the continuum limit.
In fig.\ \ref{DSL_Wilsonloops_NC}
we plot the expectation value of the Wilson loop 
in the NC plane
with the {\em same} sets of parameters
as the ones used to obtain fig.\ \ref{DSL_Wilsonloops_com}.
We do observe a compelling scaling behavior.
We have also confirmed the scaling of other quantities
\cite{Bietenholz:2006cz}.

\section{Dispersion relation}
\label{sec:dispersion-rel}

In this section we discuss the dispersion relation
in the symmetric phase.
As we mentioned before, in the present
setup we regard one of the commutative coordinates (say, $x_4$) 
as the Euclidean time.
From the exponential decay of the two-point correlation function 
of open Wilson line operators separated in the time direction,
we can extract the energy of a state that couples to this operator.
Similar studies have been done also
in the case of NC scalar field theory \cite{Bietenholz:2004xs}.

Let us define 
the open Wilson line operator at a fixed time $x_4$ as
\beq
P_\mu(x_4 , n) \equiv  \frac{1}{NL} 
\sum_{x_3} \tr \Bigl( V_\mu(x_3,x_4) ^n \Bigr) 
\nonumber
\eeq
for $\mu=1,2$, which has a zero momentum component 
in the $x_3$ direction
and a non-zero momentum component 
(\ref{momentum-polyakov}) in a NC direction
depending on $n$.
Then we define the two-point
correlation function of the open Wilson lines
\beq
C_n (\tau) \equiv \frac{1}{2} \sum_{\mu=1}^2 \sum_{x_4}
\Bigl\langle P_\mu(x_4 , n)^{*} \cdot 
 P_\mu(x_4 + \tau , n)  \Bigr\rangle
\label{2-pt-cor}
\eeq
with a separation $\tau$ in the temporal direction.

\begin{wrapfigure}{r}{6.6cm}
\centerline{\includegraphics[width=5cm,angle=270]
{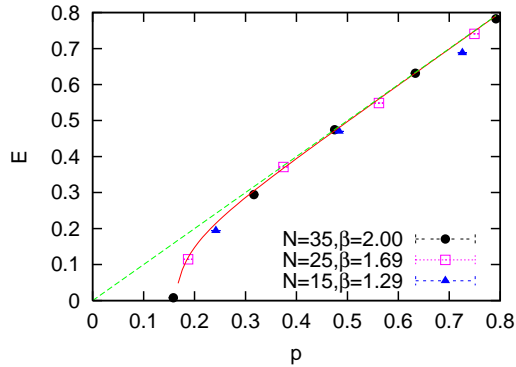}}
\caption{The dispersion relation in the symmetric phase.
The energy $E$ obtained from the 
two-point correlation function (\ref{2-pt-cor})
is plotted against the momentum $p$.}
\label{eps:dispersion}
\end{wrapfigure}

Fig.\ \ref{eps:dispersion} shows the dispersion relation 
obtained from the 
two-point correlation function (\ref{2-pt-cor}) for even $n$.
The parameters $(N,\beta )$ are chosen
as in the broken phase.
Namely, $a$ is still determined through (\ref{theta-def})
with $\theta=\frac{45}{\pi}\simeq 14.3$, and 
we fine-tune $\beta$ at each $N$ in such a way that
the scaling behavior of the Wilson loops 
in the commutative plane is optimized.
It turns out that the data points $(E,p)$ for different $N$
lie to a good approximation on a single curve
\beq
E^2= p^2 - \frac{c}{(\theta p)^2}
\label{NCdispertionat1loop}
\eeq
with $c\simeq 0.1285$. Due to the negative sign
of the second term in (\ref{NCdispertionat1loop}),
the usual Lorentz invariant (massless)
dispersion relation is bent down.
The IR singularity is regularized on the finite lattice
since the smallest non-zero momentum 
(which corresponds to $n=2$) is given
by $\frac{2\pi}{Na}\propto \frac{1}{\sqrt{N}}$.
However, if one increases $N$ at fixed $\theta$,
the energy at the smallest non-zero momentum vanishes at some $N$,
and one enters the broken phase.
Therefore we cannot take the continuum limit in 
the symmetric phase.
We have also studied the dispersion relation in the {\em broken} phase,
which reveals the existence of a Nambu-Goldstone mode associated with the 
SSB of the U(1)$^2$ symmetry.\cite{Bietenholz:2006cz}

\section{Summary and discussion}
\label{sec:summary}

To summarize, we studied four-dimensional gauge theory
in NC geometry from first principles
based on its lattice formulation.
In particular we clarified the fate of the tachyonic
instability encountered in perturbative calculations.
The lattice formulation is suited for such
a study since the IR singularity responsible for the
instability is regularized in a gauge invariant manner,
and we can trace the behavior of the system as the regularization
is removed. This revealed the existence of a first order phase
transition associated with the spontaneous breakdown of 
the U(1)$^2$ symmetry, which corresponds to the translational symmetry
in the NC directions.

The dynamical extent in the NC directions --- defined
through the eigenvalue distribution of matrices --- turns out
to be finite in the continuum limit.\cite{Bietenholz:2006cz}
An analogous first order phase transition
is found in gauge theories on fuzzy manifolds \cite{ABNN}, where
the fuzzy manifolds collapse at sufficiently large couplings.
The instability in those cases is due to the uniform condensation 
of a scalar field on the fuzzy manifold.
The phenomenon that the space-time itself becomes a dynamical object
is characteristic to gauge theories on NC geometry.
This may be related to the dynamical compactification of extra 
dimensions in string theory \cite{SSB}.
See also ref.\ \citen{Aschieri:2006uw},
which uses fuzzy spheres for compactified dimensions.

On the other hand, if we wish to obtain a 
phenomenologically viable {\em 4d model},
we may stay in the symmetric phase
by keeping the UV cutoff finite and view the NC gauge theory
as an effective theory of a more fundamental theory.
A $\theta$-deformed dispersion relation
for the photon such as the one displayed in fig.\ \ref{eps:dispersion}
should then have implications
on observational data from blazars (highly active galactic nuclei),
which are assumed to emit
bursts of photons simultaneously, covering a broad range of energy.
In particular, a relative delay of these photons 
depending on the frequency could hint at 
the existence of a NC geometry \cite{Camel}.

\section*{Acknowledgements}
J.N.\ would like to thank the organizers of the
21st Nishinomiya-Yukawa Memorial Symposium on Theoretical Physics.
Our large $N$ simulations were performed on the IBM p690 clusters of 
the ``Norddeutscher Verbund f\"{u}r Hoch- und 
H\"{o}chstleistungs\-rech\-nen'' (HLRN).

\end{document}